\documentclass[twocolumn,showpacs,preprintnumbers,superscriptaddress,amsmath,floatfix,amssymb,secnumarabic]{revtex4}

\usepackage[colorlinks=true]{hyperref}
\usepackage{graphicx}

\usepackage{caption}
\usepackage{subcaption}
\newcommand{\comment}[1]{}

\newcommand{\ev}{\, {\rm eV}}

\newcommand{\lr}[1]{ \left( #1 \right) }
\newcommand{\lrs}[1]{ \left[ #1 \right] }

\newcommand{\vev}[1]{ \langle \, #1 \, \rangle }

\newcommand{\Tr}{ {\rm Tr} \, }
\newcommand{\tr}{ {\rm Tr} \, }

\renewcommand{\det}[1]{ {\rm det} \left( #1 \right) }

\begin{document}
\sloppy

\title{Magnetism and interaction-induced gap opening in graphene with vacancies or hydrogen adatoms: Quantum Monte Carlo study}

\author{M. V. Ulybyshev}
\email{Maksim.Ulybyshev@physik.uni-regensburg.de}
\affiliation{Institute of Theoretical Physics, University of Regensburg,
D-93053 Germany, Regensburg, Universitatsstrasse 31}
\affiliation{Institute for Theoretical Problems of Microphysics, Moscow State University, Moscow, 119899 Russia}
\affiliation{ITEP, B. Cheremushkinskaya str. 25, Moscow, 117218 Russia}

\author{M. I. Katsnelson}
\email{M.Katsnelson@science.ru.nl}
 \affiliation{Radboud University, Institute for Molecules and Materials, Heyndaalseweg 135, NL-6525AJ Nijmegen, The Netherlands}
 \affiliation{Theoretical Physics and Applied Mathematics Department, Ural Federal University, Mira str. 19, Ekaterinburg, 620002 Russia}

 \date{May 1, 2015}

\begin{abstract}
We study electronic properties of graphene with finite concentration of vacancies or other resonant scatterers by a straightforward lattice Quantum Monte Carlo calculations. Taking into account realistic long-range Coulomb interaction we calculate distribution of spin density associated to midgap states and demonstrate antiferromagnetic ordering. Energy gaps are open due to the interaction effects, both in the bare graphene spectrum and in the vacancy/impurity bands. In the case of 5 \% concentration of resonant scatterers the latter gap is estimated as 0.7 eV and 1.1 eV for graphene on boron nitride and freely suspended graphene, respectively.

\end{abstract}
\maketitle

Defects effect enormously on electronic properties of graphene and other Dirac materials. In particular, vacancies in graphene are known to create mid-gap states \cite{pgc2006,katsnelsonbook} which, together with electron-electron interaction, can result in appearance of magnetic moments and rich many-body phenomena (see the review of early works in Refs.\cite{katsnelsonbook,ketal2012} and recent experimental and theoretical papers \cite{netal2012,netal2013,hetal2011,mf2013}). Note that hydrogen adatoms and some univalent organic admolecules (resonant scattering centers) produce a very similar electronic structure \cite{Wehling:10:1}. Qualitatively, it is explained by the fact that sp$^3$ state of carbon atom originated from its bond with a univalent adatom or admolecule makes it unavailable for $p_z$ electrons ($\pi$-orbitals) at energies at the neutrality point plus minus several electronvolts; for these electrons such atom is just cut from the lattice. Thus, discussing ``vacancies'' we will keep in mind also these cases; moreover, they are even closer to a simple model of vacancy just as a missed site in the honeycomb lattice (the model which will be used in our calculations) since the real vacancy produces very strong lattice distortions effecting essentially on the electronic structure \cite{Yazyev:07:1}. The case of finite concentration of vacancies is quite complicated even at a single-particle level \cite{oetal2010,yetal2012,hetal2014}. Here we consider this case taking into account a realistic model of Coulomb interaction in graphene \cite{Wehling:11:1} via the straightforward lattice quantum Monte Carlo (QMC) simulations. Keeping in mind what was said above, the best and easiest experimental realization would be partially hydrogenated graphene. We will study the antiferromagnetic (AFM) phase transition driven by the presence of adatoms. It manifests itself in emergent magnetic moments concentrated near the adatoms as well as in the band gap opening. Both features of this phase transition are very important.  Band gap, which is controllable via hydrogenation, offers an interesting possibility for prospective graphene applications in semiconductor devices. Emergent magnetic moments could be one of the reasons for short spin relaxation time in graphene, which is an essential obstacle for producing efficient spintronic devices \cite{Han2014,setal2015}. We emphasize that unlike the previous density functional theory study of vacancies in graphene \cite{Yazyev:07:1, wang2012},  here we treat very large sample  with random (non-regular) distribution of vacancies using unbiased QMC. Thus we could estimate how close vacancies influence each other in various geometrical configurations (for example, by calculation of variations of magnetic moment).  Also we could extract energy of midgap state associated with any particular scatterer and observe variations of these energies from one scatterer to another depending on its surroundings. These measurements give us an opportunity to estimate realistic width of midgap energy band for a fixed distribution of scatterers.  The demonstrated possibility of QMC simulations of large samples with arbitrary positions of scatterers is even more important because in real samples adatoms can migrate and form clusters.  Now this phenomenon is in focus of intensive research \cite{Shytov2009,garguilo2014}.  Our technique can be applied for these real spatial configurations of scatterers.  In principle, one can calculate the potential of interaction between adatoms within the QMC and to model the formation of clusters which is an interesting project for future. Here we restrict ourselves only by the case of random distribution of defects.

We start from the tight-binding Hamiltonian for noninteracting electrons with staggered mass term  which is essential in our simulations for the following reasons: (1) it eliminates zero mode in spectrum of quasiparticles thus making the fermionic operator $M$ (see below) invertible; (2) it serves as a seed for antiferromagnetic phase transition which we will study. The initial Hamiltonian without interaction and adatoms reads:
\begin{eqnarray}
\label{tbHam1}
 \hat{H}_{tb} = - t \sum\limits_{<x,y>} \lr{ \hat{a}^{\dag}_{y, \uparrow} \hat{a}_{x, \uparrow}
+  \hat{a}^{\dag}_{y, \downarrow} \hat{a}_{x, \downarrow} + h.c.}
 \pm \nonumber \\ \pm
 \sum\limits_{x} m (\hat{a}^{\dag}_{x, \uparrow} \hat{a}_{x, \uparrow} - \hat{a}^{\dag}_{x, \downarrow} \hat{a}_{x, \downarrow} ),
\end{eqnarray}
where $t = 2.7 \ev$, the sum $\sum\limits_{<x,y>}$ goes over all pairs of nearest-neighbor sites of the graphene honeycomb lattice (we impose periodic spatial boundary conditions as in Refs. \cite{Buividovich:12:1, Ulybyshev:13:1}) and the mass term has different sign at different sublattices. Here $\hat{a}^{\dag}_{x, \uparrow}$, $\hat{a}_{x, \uparrow}$ and $\hat{a}^{\dag}_{x, \downarrow}$, $\hat{a}_{x, \downarrow}$ are the creation/annihilation operators for spin up and spin down electrons at $\pi$-orbitals.

Next, we introduce the electrostatic interaction with potentials $V_{xy}$:
$ \hat{H}_{C} = {1 \over 2} \, \sum\limits_{x,y} V_{xy} \hat{q}_x \hat{q}_y$ ,
where $\hat{q}_x = \hat{a}^{\dag}_{x, \uparrow} \hat{a}_{x, \uparrow} + \hat{a}^{\dag}_{x, \downarrow} \hat{a}_{x, \downarrow} - 1$ is the operator of electric charge at lattice site $x$. The whole matrix $V_{xy}$ is constructed in the following way: at small distances (on-site interaction ($V_{xx} \equiv V_{00}$) and interactions with the nearest ($V_{01}$), next-to-nearest ($V_{02}$) and next-to-next-to-nearest ($V_{03}$) neighbours)  we use the potentials calculated by the constrained RPA method \cite{Wehling:11:1};  at larger distances we use ordinary Coulomb  $V_{xy} = V_C r_{01} / r_{xy}$. Parameter $V_C$ defines the strength of the Coulomb tail. Since $r_{01}$ is the distance between the nearest neighbours, $V_C$ is equal to the Coulomb repulsion energy at the distance of conjugated C-C bond. There is a small difference in calibration of our model ($t = 2.7 \ev$) and the calculations \cite{Wehling:11:1}, where the hopping parameter was $2.8 \ev$ but any possible changes of observable quantities due to this difference are definitely outside the accuracy of the applied QMC technique.

We use two settings of interaction potentials. The first one is called ``ordinary potentials'' and corresponds to freely suspended graphene (the dielectric constant of environment is equal to one). At small distances it is simply the set of potentials from the paper \cite{Wehling:11:1}: $V_{00}=9.3$ eV, $V_{01}=5.5$ eV, $V_{02}=4.1$ eV and $V_{03}=3.6$ eV. The strength of Coulomb tail is defined by the ratio  $V_{03} = V_C r_{01}/ r_{03}$, thus $V_C=7.2$ eV. It is a simple continuous extension of the potentials at small distances. The second setting is specified as ``screened potentials'' and corresponds to graphene at a substrate with the dielectric constant $\epsilon=4.5$ which is roughly the value reasonable for both boron nitride and SiO$_2$. In this case the Coulomb tail is $(1+\epsilon)/2=2.75$ times weaker while $V_{02}$  and $V_{03}$ are 1.5 times smaller. $V_{00}$ and $V_{01}$ are untouched since the screening by substrate should be irrelevant at a few interatomic distances.

 As was already mentioned, the vacancy or adatom is modeled by setting to zero all hoppings to the vacant site. We also exclude the vacant sites from the interaction term $H_C$ because they have constant zero charge.
\begin{figure}
  \centering
		\includegraphics[width=.45\textwidth]{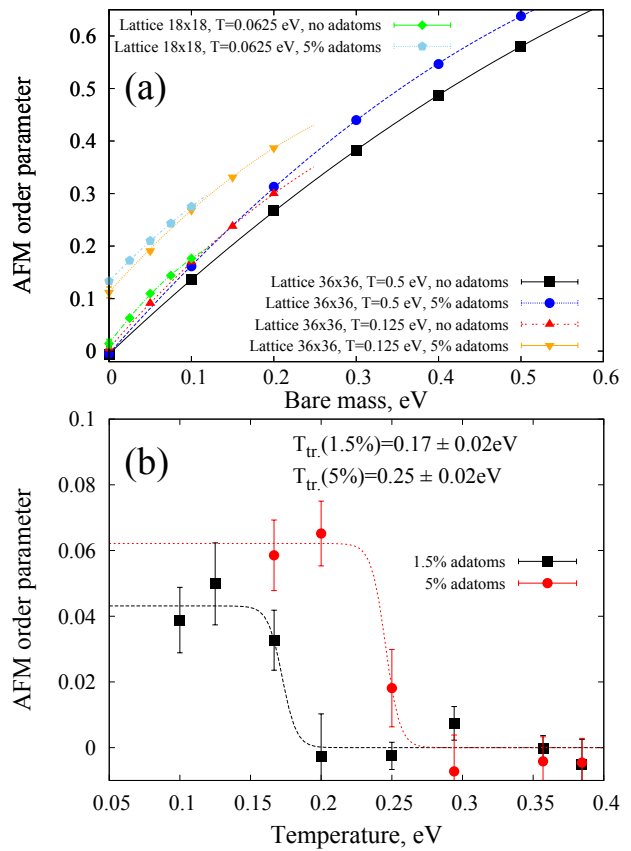}
	\caption{(a) Antiferromagnetic order parameter at various temperatures, calculated on different lattices in presence of adatoms and without them. (b) Temperature dependence of  antiferromagnetic order parameter in the zero bare mass limit. Calculation was performed on the $12 \times 12$ lattice.}
 \label{fig:condensate}
   \end{figure}
We employ Hybrid Monte-Carlo algorithm, broadly used in lattice QCD. It was applied earlier for studies of graphene in the papers \cite{drut2009, armour2010}, where so-called staggered fermions were used to model low-energy effective field theory of graphene. This algorithm was developed further in Refs. \cite{Rebbi:11:1, Buividovich:12:1, Ulybyshev:13:1}. At first, we perform Suzuki-Trotter decomposition of partition function $\exp (-\beta H)$ in order to represent it as a functional integral over trajectories in Euclidean time. In order to get rid of four-fermionic terms in the Hamiltonian, we use Hubbard-Stratonovich transformation and obtain the following partition function after integrating out fermionic fields:
\begin{eqnarray}
\label{PartFunc2}
 \tr e^{-\beta \hat{H}} \cong \int \mathcal{D}\varphi_{x,n} e^{-S\lrs{\varphi_{x,n}}}
 |\det{M\lrs{\varphi_{x,n}}}|^2,
\end{eqnarray}
where $\varphi_{x,n}$ is the Hubbard-Stratonovich field for timeslice $n$ and spatial coordinate $x$. $\delta \tau N_t=\beta$, where $\delta \tau$ is the step in Euclidean time, $N_t$ is number of steps and $\beta$ is inverse temperature, $M$ is the fermionic operator (inverse fermionic Green's function at a given configuration of auxiliary field). We use its particular form \cite{Ulybyshev:13:1}. In more details (including issues with continuous limit $\delta \tau \rightarrow 0$) it was discussed in Ref. \cite{Smekal:14:1}. The particle-hole symmetry for graphene at neutrality point makes the integration weight in (\ref{PartFunc2}) positive due to appearance of the squared modulus of the determinant, thus, we have no fermionic sign problem \cite{Raedt1985}.  For both sets of inter-electron interaction potentials, the action of the Hubbard-Stratonovich field $S\lrs{\varphi_{x,n}}$ is also a positive definite quadratic form. Thus we generate configurations of $\varphi_{x,n}$ by a Monte-Carlo method and calculate physical quantities as averages over the generated configurations. Here we follow Refs. \cite{Rebbi:11:1, Buividovich:12:1, Ulybyshev:13:1} and use so-called $\Phi$-algorithm.

We used lattices with spatial sizes $18 \times 18$, $24 \times 24$ and $36 \times 36$ in order to study finite-size effects. We studied lattice with 5 \% adatoms (in the most of calculations), scattered uniformly in the whole sample. Three different temperatures were studied: T=0.5 eV (corresponds to $N_t=20$), T=0.125 eV ($N_t=80$) and T=0.0625 eV ($N_t=160$). For all temperatures we generated configurations with four masses, for example in the case of T=0.125 eV we used $m=0.05, 0.1, 0.15, 0.2$ eV. Physical results are obtained via extrapolation to zero mass. In all calculations except energies of midgap states we use ``ordinary potentials''.

\begin{figure}
  \centering
		\includegraphics[width=.45\textwidth]{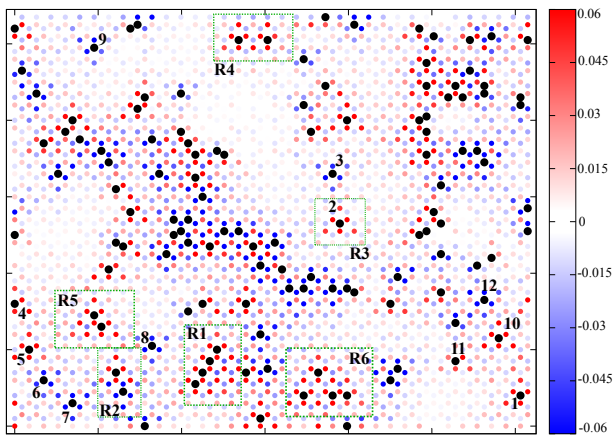}
	\caption{Distibution of average spin. Color scale corresponds to $\langle S_z \rangle$ at the site in the zero bare mass limit.}
 \label{fig:condensate}
   \end{figure}

According to Lieb theorem for the Hubbard model \cite{Lieb:89:1,katsnelsonbook} the ground state for the case of vacancies equally distributed between two sublattices should be spin singlet, and there are no physical reasons to expect that the long-range character of Coulomb interactions can change this conclusion. Keeping in mind that single vacancy or adatom induces magnetic moment one should consider opportunity of antiferromagnetic ordering at finite concentration (ferromagnetism is impossible). In this case the order parameter is the difference in average spin between sublattices (denoted as $A$ and $B$ in the formula):
$ \vev{ \Delta n } =\vev{ {1 \over N_A} \sum\limits_{x \in A} (\hat{a}^{\dag}_{x, \uparrow} \hat{a}_{x, \uparrow} -  \hat{a}^{\dag}_{x, \downarrow} \hat{a}_{x, \downarrow}) -
{1 \over N_B} \sum\limits_{x \in B} (\hat{a}^{\dag}_{x, \uparrow} \hat{a}_{x, \uparrow} - \hat{a}^{\dag}_{x, \downarrow} \hat{a}_{x, \downarrow}) }$,
$N_A$ and $N_B$ are the overall number of sites in A and B sublattice, respectively.  The results are presented in Fig. \textcolor{red}{1a}. In case of the highest temperature (0.5 eV) the order parameter is equal to zero in the physical limit of zero bare mass disregarding the presence of adatoms. Only at lower temperature (0.125 eV) the order parameter acquires nonzero value in presence of adatoms and remain almost stable with further decreasing of the temperature (0.0625 eV).   Fig. \textcolor{red}{1b} presents the temperature dependence of AFM order parameter. This calculation was also performed using one particular random distribution of adatoms for each concentration. One can clearly see a sharp transition at certain critical temperature (Neel temperature).  The results were fitted with ``step function'' $f(T)=C ( 1- \tanh ( b( T-a)))$, where parameter $a$ gives us the value of the critical temperature.  Thus we have an estimation of effective antiferromagnetic coupling (for 1 \% of defects) $J \sim 0.1$ eV.  This value is two orders of magnitude larger than the one estimated from recent experimental data for vacancies in graphene \cite{morgen2014}. This is an important point showing that probably exchange interaction is very sensitive to real electronic srtucture (we mentioned in the introduction that for the real vacancies it is very strongly affected by atomic reconstruction) and that the use of the simplest one-band tight-binding model instead of full-electron calculations can be dangerous for the problems related to magnetism. Recent density functional calculations of exchange interactions in single-site hydrogenated or fluorinated graphene \cite{Rudenko2013} predict complicated noncollinear magnetic ground states, in a sharp contrast with predictions of Lieb theorem for the single-band model (saturated ferromagnetism). This issue requires further investigations. 

Spatial distribution of electron spin density is presented in Fig. \textcolor{red} {2}. It represents the quantity $f_x=\langle {\hat{a}^\dag}_{x, \uparrow} \hat{a}_{x, \uparrow} \rangle$ at each lattice site. Since the particle-hole symmetry is unbroken, the equality $\langle {\hat{a}^\dag}_{x, \uparrow} \hat{a}_{x, \uparrow} \rangle + \langle {\hat{a}^\dag}_{x, \downarrow} \hat{a}_{x, \downarrow} \rangle =1$ is satisfied exactly  for each lattice site. It means that regions with positive $f_x$ have non-compensated spin up and negative $f_x$ corresponds to non-compensated spin down. It is clearly seen that antiferromagnetic order is generated in the vicinity of adatoms. Moreover, one isolated adatom has nonzero average spin (see the first row in the table \ref{tab:spin}). This spins tend to be parallel for adatoms at one sublattice and antiparallel for adatoms at different sublattices. If adatoms are placed equivalently at both sublattices, they generate the same spin excess at both sublattices and thus the full spin will be close to zero. This means that, indeed, the statement of the Lieb theorem \cite{Lieb:89:1} remains correct in the case of long-range Coulomb interaction. More detailed description of spin-spin correlations including explicit evidence of spontaneous breaking of SU(2) spin symmetry is presented in supplementary material.

\begin{table}[t]
  \begin{tabular}{ | c | c |}
    \hline
    Region on the map & $M= 2 \langle S_z \rangle$ \\  \hline  \hline
    R3 & 0.530 $\pm$ 0.016 \\ \hline
    R4 & 1.330 $\pm$ 0.026 \\ \hline
    R5 & 1.542 $\pm$ 0.026 \\ \hline
    R6 & 2.70 $\pm$ 0.04 \\ \hline
    R1 & 3.20 $\pm$ 0.04 \\
    \hline
  \end{tabular}
\caption{Average magnetic moment of different configurations of adatoms. R3 corresponds to one isolated adatom, R4 corresponds to 2 adatoms at the distance of 2 lattice steps, R5 contains 2 adatoms at the distance of 1 lattice step, R1 and R6 contain 4 adatoms (denser configuration in case of R1).}
  \label{tab:spin}
\end{table}
In order to characterize the correlation between adatoms at one sublattice quantitatively, we have measured full magnetic moment $M=2 \mu_B \langle S_z \rangle$ of some spatial configurations of vacancies.  Results (in units of Bohr's magneton) are summarized in the table \ref{tab:spin}. One can observe strong dependence of magnetic moment on the geometry of the adatoms configurations. For example, two adatoms at the distance of 1 lattice step have magnetic moment 1.5 times larger than two isolated adatoms.
\begin{figure}
  \centering
		\includegraphics[width=.45\textwidth]{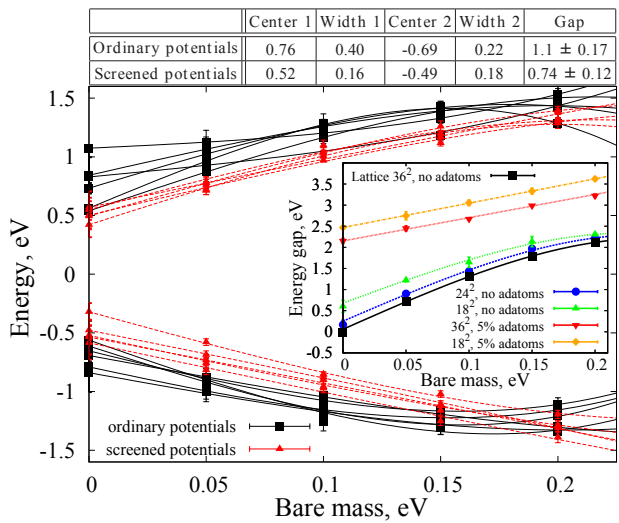}
	\caption{Energies of midgap states for two sets of inter-electron potentials. Each state correspond to one isolated vacancy marked with a number in the figure \textcolor{red} {2}. Center and width of the bands are calculated in the limit $m \rightarrow 0$. Center is average over energies of all states in each band, width is equal to doubled dispersion.  T=0.125 eV. Real physical situation is restored in the limit $m \rightarrow 0$. (inset) Energy gap between "normal" energy bands.  All values correspond to the K-point in Brillouine zone. }
   \label{fig:energies}
  \end{figure}

The second set of calculations is devoted to measurements of mass gap in the presence of vacancies/adatoms. We study two types of energy bands: ``normal'' energy bands which transfer into Dirac cones in the absence of adatoms and midgap states which are concentrated in the vicinity of isolated adatoms. In the latter case we perform calculations for both sets of potentials to measure the influence of screening on the energies of midgap states. Calculation of the energies is based on the two-point Euclidean Green functions contracted with some guess for projector to a wavefunction $\psi(x)$ of the state we are interested in:
\begin{eqnarray}
\label{green_function}
 C (\tau) = \sum\limits_{x, y} \Tr \left({ \hat{a}^{\dag}_{x} \bar \psi(x) e^{-\tau \hat H} \hat{a}_{y} \psi(y) e^{- (\beta-\tau) \hat H} }\right).
\end{eqnarray}
At large enough $\tau$ this correlator is proportional to $e^{-\tau E_0}$ where $E_0$ is energy of the state under study. In the case of ``normal'' energy band we use lattice exponent $\exp{(i \vec k \vec x)}$ concentrated at one sublattice with wave vector $\vec k$ at the K-point of Brillouin zone as a guess for the wavefunction. Therefore we are able to estimate the lower bond of the energy band and the energy gap between these bands. In the case of midgap state we guess that wavefunction is concentrated in the three nearest neighbours of the vacant site. In order to check these measurements we perform the same calculation for freely suspended graphene without vacancies. In this case the gap should be equal to zero in the zero bare mass limit \cite{Ulybyshev:13:1}. Results for the ``normal'' energy band are presented in the inset on Fig. \ref{fig:energies}. In presence of vacancies we use simple linear fit. Without vacancies  the polynomial fit $\phi(m)=c_0+c_1 m + c_2 m^2$ is employed.  For the largest lattice ($36 \times 36$) $c_0$ is zero within the errorbars, so the fitting works well and this lattice is large enough to reproduce zero gap in the $m_{bare} \rightarrow 0$ limit. For smaller lattices one can observe non zero $c_0$ due to large  finite-size effects.

For midgap states we used wavefunctions concentrated near 12 relatively isolated adatoms, marked with black numbers in the figure \textcolor{red} {2}.  Summarizing all these calculations (see Fig. \ref{fig:energies}) we conclude that states concentrated near adatoms form two rather broad bands in between of ``normal'' energy bands. Positive and negative energies of midgap states correspond to adatoms at different sublattices. The gap between these bands is calculated as a distance between two levels with the smallest absolute values of energies. It can exceed 1 eV for suspended graphene, but decreases for graphene at substrate. Width of the bands is a measure of the interplay between midgap states concentrated near different vacancies. Obviously, if concentration of adatoms tends to zero, the energies of the midgap  states will be almost constant. The same effect is observed here in case of suppressed Coulomb tail: midgap states near isolated vacancies recognize their surrounding more poorly.

To conclude, electron-electron interactions for finite concentration of adatoms lead to antiferromagnetic ordering, in a qualitative agreement with Lieb theorem despite its formal inapplicability to systems with long-range Coulomb interactions. Even probably more interestingly they result in gap opening: a ``big gap'' of the order of several eV at the K point and ``smaller gap'' (but still quite noticeable, about 1 eV for 5\% concentration of adatoms) in the mid gap states. The latter prediction can be checked by measuring optics for chemically functionalized graphene. One could expect that the effects of disorder will smear the gap in the defect band transforming it into a pseudogap. One could hope however that the two-peak structure characteristic of the pseudogapped state can be distinguished from a single-peak structure centered new the neutrality point predicted by the single-particle picture \cite{Yuan2010}. Alternatively, the reconstruction of the defect band can be studied experimentally by scanning tunneling spectroscopy. 

\begin{acknowledgments}
We thank Dr. P. V. Buividovich for useful discussions. The work of MU was supported by the DFG grant SFB/TR-55 and by Grant RFBR-14-02-01261-a. Numerical calculations were performed at the ITEP computer systems ``Graphyn'' and ``Stakan'' and at the supercomputer center of Moscow State University. MIK acknowledges funding from the European Union Seventh Framework Programme under Grant Agreement No. 604391 Graphene Flagship and from NWO via Spinoza Prize.
\end{acknowledgments}

\newpage
\begin{flushleft}
\textbf{ \Large{ Supplemental material}}
\end{flushleft}

\renewcommand{\thefigure}{S\arabic{figure}}

\setcounter{figure}{0}

\begin{figure}[b]
  \centering
		\includegraphics[width=0.3\textwidth, angle=270]{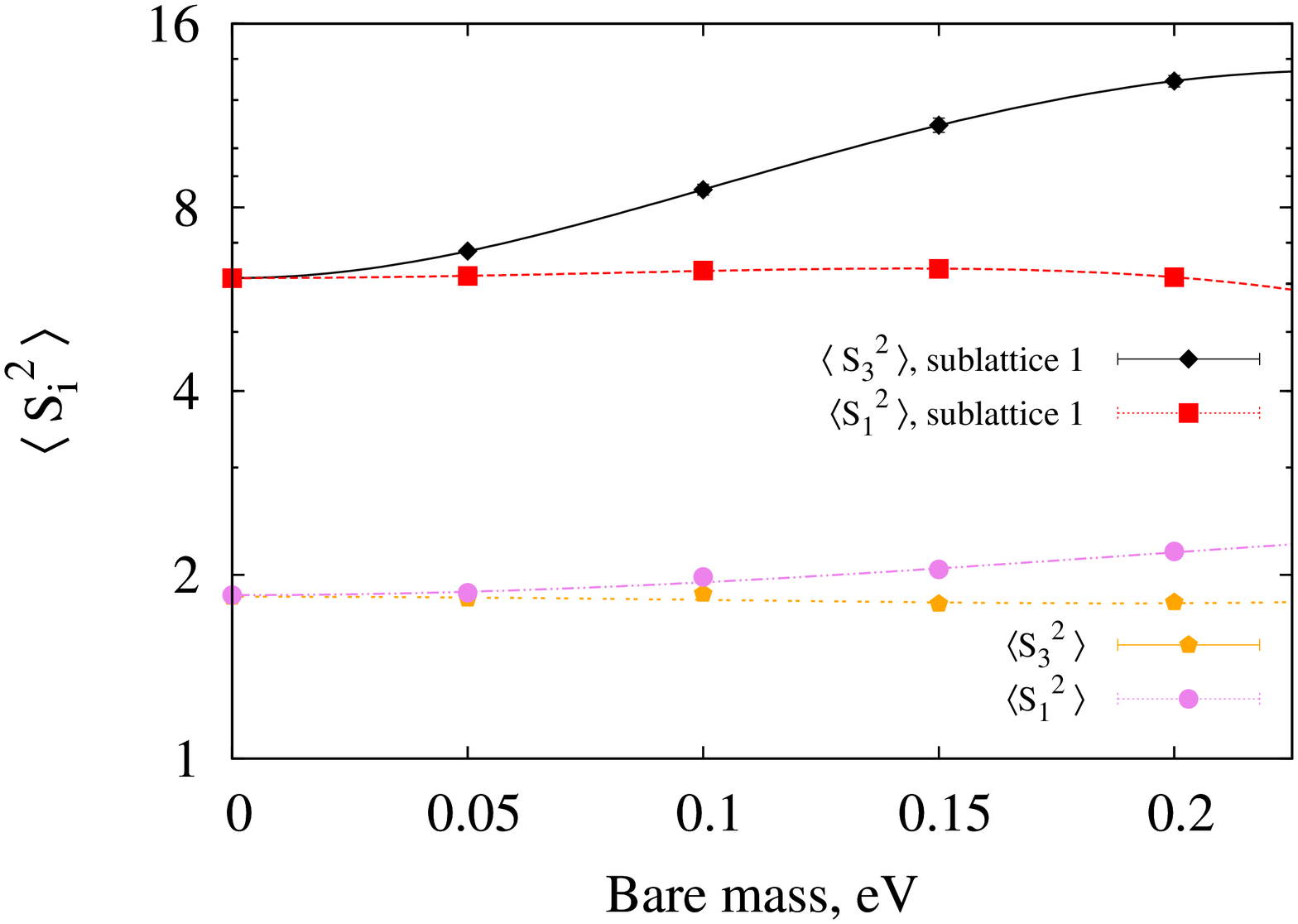}
	\caption{Dependence of $\langle {S_i}^2 \rangle$ on bare mass in absence of adatoms. Lattice size is $36 \times 36$, T=0.125 eV.  $S$ is full spin of a 6 × 6 cells region of the lattice. Data for the second sublattice is not shown since it is exactly identical to the first one. }
  \label{fig:spin_m}
\end{figure}

\begin{figure}[h]
  \centering
		\includegraphics[width=0.3\textwidth,  angle=270]{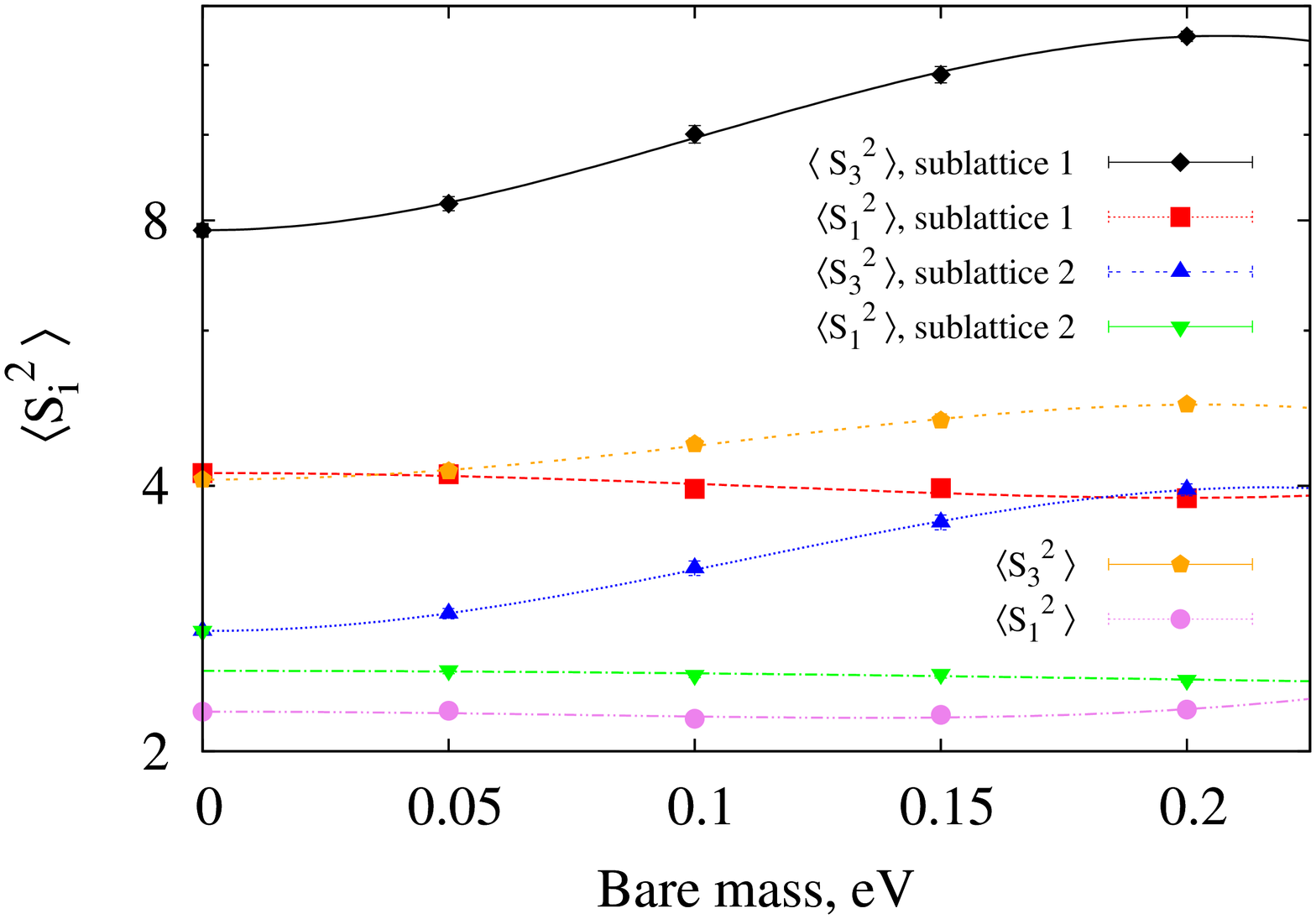}
	\caption{Dependence of $\langle {S_i}^2 \rangle$ on bare mass in presence of 5\% adatoms. Lattice size is $36 \times 36$, T=0.125 eV.  $S$ is full spin of the region R1 (see fig. \textcolor{red} {2} in the main text). }
  \label{fig:spin_m1}
\end{figure}

More close view at average spin is presented in the figures \ref{fig:spin_m} and \ref{fig:spin_m1}. They demonstrate dependence of $\langle {S_{3}}^2 \rangle$ and $\langle {S_{1}}^2 \rangle$ on bare mass in presence of vacancies (\ref{fig:spin_m1}) and without them (\ref{fig:spin_m}). $S$ is a full spin for some region of the lattice: $S_i=\sum_{x \in R} S_{x,i}$, where $S_{x,i} = 1/2 ({\hat{a}^{\dag}}_{x, \uparrow}, \quad {\hat{a}^{\dag}}_{x, \downarrow}) \sigma_i \left( {{\hat{a}}_{x, \uparrow}} \atop {\hat{a}_{x, \downarrow}} \right)$; $\sigma_i$ are Pauli matrices.  $S^{(\alpha)}_i$ is a spin for particular sublattice $\alpha=1,2$. Since $\langle {S^{(\alpha)}_i}^2 \rangle$ are even in bare mass, we use the following  polynomial extrapolation: $\phi(m)=c_0+c_1 m^2 + c_2 m^4$. It is clearly seen that both sublattice and SU(2) spin symmetry are restored in the limit of zero mass in absence of vacancies ($\langle {S^{(\alpha)}_{1}}^2 \rangle = \langle {S^{(\alpha)}_{3}}^2 \rangle$ at both sublattices in the limit $m \rightarrow 0$, see \ref{fig:spin_m}). Another situation can be observed in presence of adatoms at \ref{fig:spin_m1}. We show $\langle {S^{(\alpha)}_{i}}^2 \rangle$ for the lattice with 5 \% adatoms choosing the region where adatoms are concentrated mostly at one sublattice (region R1 in the figure \textcolor{red} {2} in the main text). It is clear that both sublattice and SU(2) spin symmetry are broken: $\langle {S^{(\alpha)}_{1}}^2 \rangle \neq  \langle {S^{(\alpha)}_{3}}^2 \rangle$ at both sublattices  even in the zero mass limit.

\begin{figure}
  \centering
		\includegraphics[width=.3\textwidth, angle=270]{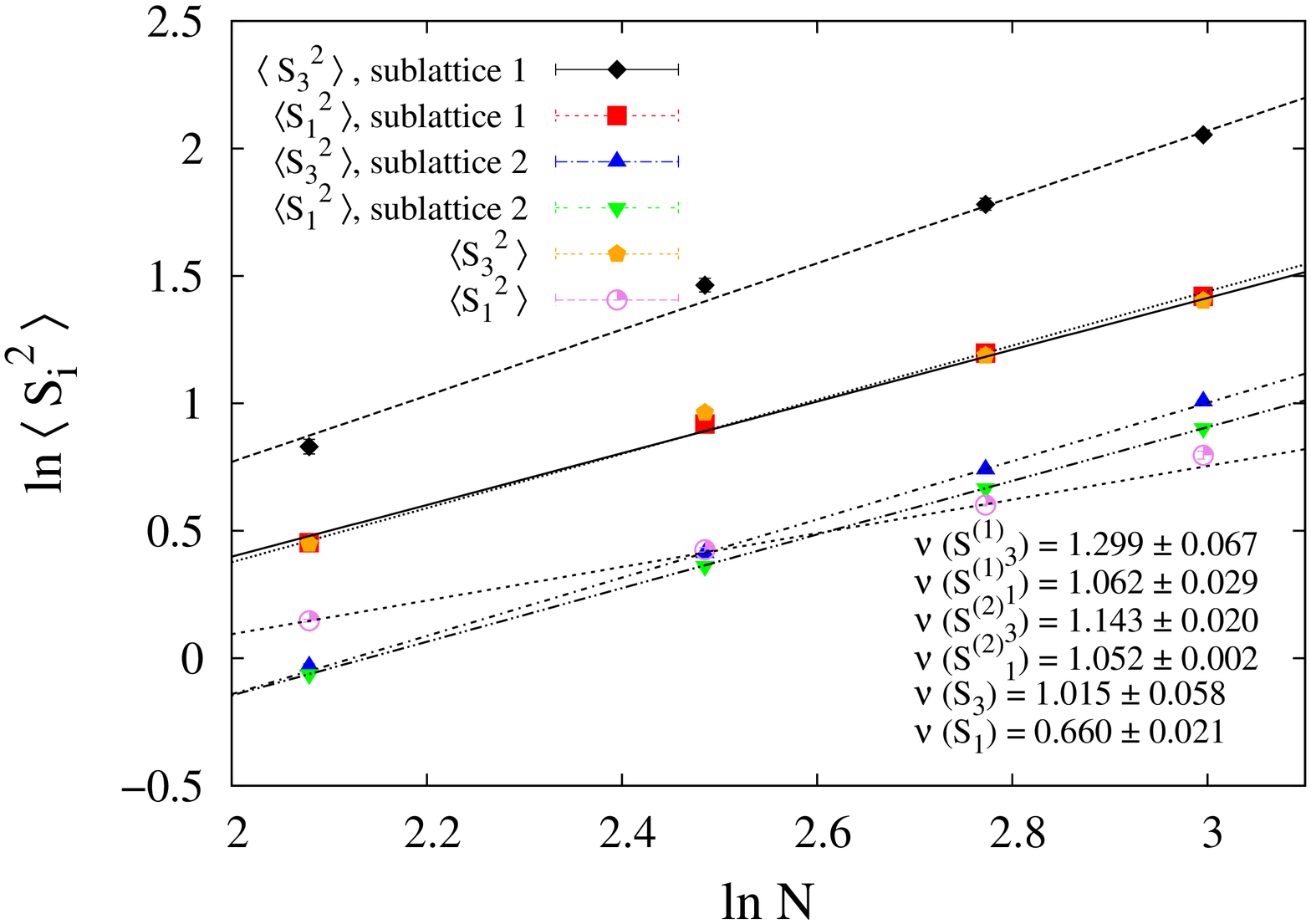} 
	\caption{Dependence of $\langle {S_i}^2 \rangle$ on the size of the system. Calculation is performed on the lattice with 5 \% adatoms inside the regions R1  (see fig. \textcolor{red} {2} in the main text). T=0.125 eV. All the results are shown in the $m \rightarrow 0$ limit.}
 \label{fig:spin_r}
   \end{figure}

\begin{figure}
  \centering
		\includegraphics[width=.3\textwidth, angle=270]{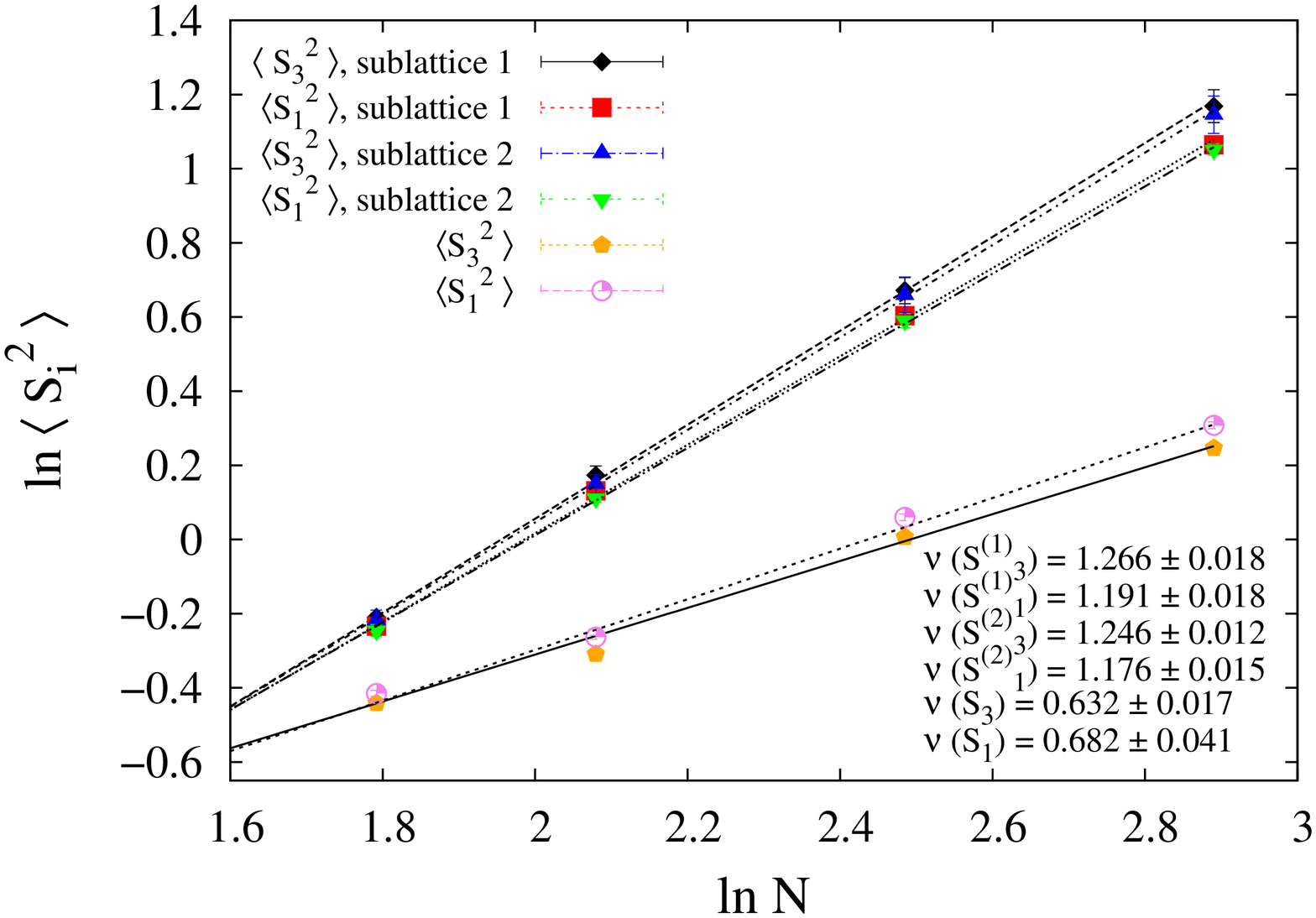}
	\caption{Dependence of $\langle {S_i}^2 \rangle$ on the size of the system. Calculation is performed on the lattice with 5 \% adatoms inside the regions R2 (see fig. \textcolor{red} {2} in the main text). T=0.125 eV. All the results are shown in the $m \rightarrow 0$ limit.}
 \label{fig:spin_r1}
   \end{figure}

Now let us turn to spin-spin correlations for different spatial configurations of adatoms. Figures \ref{fig:spin_r}  and \ref{fig:spin_r1} demonstrate the dependence of $\langle {S^{(\alpha)}_i}^2 \rangle$ on the size of the system. In the region R1 (see figure \textcolor{red} {2}) adatoms are concentrated at the 2d sublattice, while in the region R2 adatoms are placed equivalently at different sublattices. In the first case (\ref{fig:spin_r}) $S^{(\alpha)}_3$ spin components at different sites within one sublattice  are correlated with each other: ${(S^{(1,2)}_{3})}^2 \sim N^\nu$ with $\nu>1$. The strongest correlation is within the 1st sublattice (the  ``red'' one in the figure \textcolor{red} {2}) while $S^{(\alpha)}_1$ are almost uncorrelated. Again, $S^{(\alpha)}_1$ and $S^{(\alpha)}_3$ spin components behave differently because of spontaneously broken SU(2) symmetry. Since there is no difference between $S^{(\alpha)}_1$ and $S^{(\alpha)}_2$, the SU(2) symmetry is broken up to U(1) rotations. From the calculation of full spin (it includes both sublattices) we see that both $S_3$ and $S_1$ are anticorrelated at different sublattices, because $\nu$ sufficiently decreases in both cases. In the second configuration of adatoms (figure \ref{fig:spin_r1})   $S^{(\alpha)}_3$ and $S^{(\alpha)}_1$ are correlated equivalently inside one sublattice. All components of electron spin at different sublattices are again anticorrelated. Thus we really have a spontaneous breaking of SU(2) spin symmetry and sublattice $Z_2$ symmetry leading to antiferromagnetic  spin ordering. 


\end{document}